\documentclass[a4paper,11pt]{article}
\usepackage{pos}
\usepackage{subcaption}
\usepackage{svg}
\svgsetup{
          inkscapepath=svgsubdir,inkscapeformat=png
         }
\usepackage{mathtools}
\usepackage{siunitx}
\usepackage{cleveref}
\usepackage{physics}
\usepackage{braket}

\crefformat{footnote}{#2\footnotemark[#1]#3}

\DeclareMathOperator{\im}{i}
\newcommand*{\SU}[1][3]{\ensuremath{\text{SU}\qty(#1)}}

\title{Investigating vector boson scattering:\\ A fully gauge-invariant study}
\ShortTitle{Investigating vector boson scattering}

\author*{Bernd Riederer}
\emailAdd{bernd.riederer@uni-graz.at}
\author{Axel Maas}
\emailAdd{axel.maas@uni-graz.at}

\affiliation{Institute of Physics, NAWI Graz, University of Graz,\\
Universitätsplatz 5, Graz, Austria}

\abstract{
  Vector boson scattering (VBS) plays a central role in the search for new physics at collider experiments such as ATLAS and CMS at the LHC. Usually predictions for this kind of process are obtained using mainly perturbative approaches in fixed gauges. Here we present a fully gauge-invariant study of VBS in the scalar-channel involving three different types of Higgs-like particles characterized by their mass; above (heavy), inside (resonance) or below (stable) the elastic region. To this end, we combine results obtained in a reduced SM setup from (augmented) perturbation theory with those from non-perturbative lattice simulations.
}

\FullConference{%
  The 39th International Symposium on Lattice Field Theory,\\
  8th-13th August, 2022,\\
  Rheinische Friedrich-Wilhelms-Universität Bonn, Bonn, Germany\\ 
  \;\\
  and\\
  \;\\
  41st International Conference on High Energy physics - ICHEP2022\\
  6-13 July, 2022\\
  Bologna, Italy
}

\begin{document}
\maketitle

\section{Introduction}\label{sec:intro}

Vector boson scattering (VBS) has gained a lot of experimental interest in the recent years, especially as a possible path to search for physics beyond the standard model (BSM) \cite{Gallinaro:2020cte}. Within the standard model VBS can be very well described using only the electroweak sector in combination with the Brout-Englert-Higgs (BEH) effect. In a standard perturbative framework scattering quantities, like differential or integrated cross-sections as well as phase-shifts, can be reliably obtained within this subset of the SM \cite{Lee:1977eg,Denner:1996ug,Denner:1997kq,Buarque:2021dji}. Modifications to VBS from next-to-leading-order (NLO) effects of additional SM-processes, like QCD, influence these quantities by $\sim$\SI{10}{\percent} at the LHC \cite{Biedermann:2017bss,Denner:2019tmn,Denner:2020zit,Denner:2022pwc}.

However, in usual PT the elementary fields of the theory are considered as the physical degrees of freedom which cannot be the case \cite{Banks:1979fi,Frohlich:1980gj,Osterwalder:1977pc}. Since they are charged under the weak gauge group they are gauge-dependent and thus unphysical \cite{Frohlich:1980gj}. Instead, it is necessary to choose a completely gauge-invariant approach to obtain physical scattering quantities. As it turns out this requires using composite objects (i.e.\ bound states) as the elementary degrees of freedom, see for a review \cite{Maas:2017wzi}.

Here we focus on the fully gauge-invariant description of VBS for a reduced standard model setup with different types of physical Higgs-bosons: a stable Higgs-boson (like in the electroweak SM), a Higgs-like resonance\footnote[0]{\label{note:skip}This part/section has been left out for the ICHEP2022 proceedings. Important parts of \cref{sec:lattice} are collected in an additional paragraph at the beginning of \cref{sec:results}.}, and a Higgs-boson above the inelastic threshold\cref{note:skip}. We introduce the necessary framework for the gauge-invariant description in \cref{sec:higgs_radius}, namely the Fröhlich-Morchio-Strocchi-mechanism (FMS) \cite{Frohlich:1980gj} and augmented perturbation theory (APT) \cite{Maas:up,Maas:2017wzi,Maas:2020kda}. From this the expected modifications of the scattering observables due to the finite extent of the involved particles are obtained and compared to usual PT. Additionally, to the perturbative approach we also show results obtained using lattice simulations. Both approaches are then collected in \cref{sec:results} and compared with each other. This gives us finally a clear picture of VBS. Technicalities concerning the involved lattice simulations are given in \cref{sec:lattice}\cref{note:skip}. For a more detailed discussion of the methods used, and additional results, see \cite{Jenny:2022atm}.

\section{A Higgs with finite extent}\label{sec:higgs_radius}

A fully gauge-invariant approach to scattering processes requires using composite objects as the elementary degrees of freedom. The reason for this lies in the nature of the BEH construction itself. Therefore, all theories that are using this mechanism directly, or those which can effectively be described by it (e.g.\ composite Higgs), follow the same reasoning \cite{Maas:2017wzi}. To illustrate the issue let us consider a scalar field $\phi$ that is in the fundamental representation of the gauge group one is interested in. For the electroweak sector this group would be $\SU[2]_W$. This field is further restricted by a potential of the form
\begin{equation}
  V(\phi) = \lambda\qty(\phi^\dagger \phi -f^2)^2\,,
\end{equation}
which is invariant under the gauge transformation $\phi(x)\rightarrow G(x)\phi(x)$, with $G(x)$ some element of the gauge group. However, the potential has a non-trivial minimum at $\phi^\dagger \phi=f^2$. For the usual perturbative procedure of the BEH effect \cite{Bohm:2001yx,Maas:2017wzi} one needs to select a particular minimum of the potential by fixing the gauge, e.g.\ 't Hooft gauge, followed by a shift of the field according to $\phi\rightarrow v + \eta$, where $\abs{v}=f$ is the vacuum expectation value (VEV). This results in mass terms at tree-level for the gauge-bosons and is commonly called ``\emph{spontaneous gauge-symmetry breaking}''.

Consequently, the elementary fields are used to obtain cross-sections by calculating the corresponding matrix elements, see e.g.\ \cite{Bohm:2001yx}. The problem with this approach is that the shift in the BEH effect requires gauge-fixing, which indeed is the only possibility to do so. So in fact these elementary fields are gauge-dependent, which renders these states unphysical \cite{Frohlich:1980gj}. This problem is usually circumvented by applying a BRST construction \cite{Bohm:2001yx} to identify the physical degrees of freedom. However, beyond perturbation theory this construction does not hold anymore due to the existence of Gribov copies even at arbitrary weak coupling \cite{Fujikawa:1982ss,Maas:2017wzi}.

The only remaining way out is therefore to use inherently gauge-invariant operators (i.e.\ composite operators) as the physical degrees of freedom. For the process of VBS the two operators with the correct spin and parity quantum numbers $J^P$ of the Higgs- and gauge-bosons are given by:
\begin{align}\label{eqn:physical_ops}
  \mathcal{O}^{H}            = \mathcal{O}^{0^{+}}             =  \phi^{\dagger}\phi &  &
  \mathcal{O}^{W\, a}_{\mu}  = \mathcal{O}^{1^{-}\, a}_{\mu}   =  \phi^{\dagger}D_{\mu}^{a}\phi\,,
\end{align}
with $D_\mu^a$ the covariant derivative of the gauge group. Since the physical degrees of freedom are now described by these gauge-invariant operators we see that the BEH effect indeed requires them to be bound states and thus having a non-zero radius. Or, to be more precise, they are not point-like objects anymore as in standard PT.

To obtain scattering quantities for these operators it seems now unavoidable to use nonperturbative methods, like it is done in QCD. However, when considering instead the usual approach to the BEH effect this raises the question of why it agrees so strikingly well with experimental results \cite{ParticleDataGroup:2020ssz,Bohm:2001yx} while neglecting the inherently nonperturbative structure. This suggests some correspondence between the usual perturbative treatment and the fully gauge-invariant approach. 

To get a better understanding lets deal a bit more with the weak sector of the SM, which contains exactly these building blocks. Our previous arguments now suggest that instead of elementary fields the operators (\ref{eqn:physical_ops}) should be used as asymptotic states to obtain correlators, which needs nonperturbative methods in principle. However, due to the BEH-effect it is possible to augment perturbation theory by an additional step that preserves gauge-invariance, while still allowing perturbative access to the quantities of interest. The full procedure will therefore be called augmented perturbation theory (APT) and consists of two steps: the FMS-expansion \cite{Frohlich:1980gj,Maas:2017wzi} followed by usual PT.

As an example we take the propagator of a physical Higgs-boson given by $\expval{\mathcal{O}^{H}(x)^{\dagger}\mathcal{O}^{H}(y)}$. The first step of APT is to insert the usual BEH split in a convenient gauge. This leaves us with a sum of (individually gauge-dependent) correlation functions, e.g.\
\begin{equation}
  \expval{ \qty[\phi^{\dagger}\phi]\!(x)^\dagger\qty[\phi^{\dagger}\phi]\!(y)\!}=\expval{\qty[v\eta]\!(x)^\dagger \qty[v\eta]\!(y)\!}
  +\expval{\qty[\eta^\dagger\eta]\!(x)^\dagger \qty[v\eta]\!(y)+(x)\!\leftrightarrow\!(y)\!}+\expval{\qty[\eta^\dagger\eta]\!(x)^\dagger\qty[\eta^\dagger\eta]\!(y)\!}.
\end{equation}
In a second step a double expansion in $v$ and the other coupling constants can be made. At leading order in $v$, the propagator of the composite operator $\mathcal{O}^{H}$ therefore coincides to all orders in all other couplings with the elementary Higgs propagator $\expval{\eta(x)^\dagger \eta(y)}$, and especially has the same mass and width \cite{Maas:2020kda}. This procedure can be  applied to any correlation function, and also to matrix elements as will be seen in \cref{sec:results}. Due to the finite extent of the observable particles, modifications to scattering quantities from off-shell contributions are expected compared to usual PT.

So far we have motivated that the gauge-invariant description of weakly interacting particles in the SM should give them some finite radius. Although, this will not change masses and decay-widths of the particles involved in VBS it still may modify cross-sections. In addition, some BSM models, like composite Higgs, directly introduce a finite extent to the particles involved in the VBS process. Therefore, regardless of the previous motivations, it is worthwhile to consider here what modifications are to be expected from a non-vanishing radius of the involved physical particles.

In principle to get measurable predictions it is needed to obtain (differential) cross-sections from fundamental theory. Therefore, experiment and theory can be connected via the equation
\begin{equation}
  \dv{\sigma}{\Omega} = \frac{1}{64\pi^2 s}\abs{\mathcal{M}}^2
\end{equation}
with $\dv{\sigma}{\Omega}$ the differential cross-section, $\sqrt{s}$ the center of mass energy and $\mathcal{M}$ the transition matrix. This matrix can be obtained for any theory and process from the possible exchange diagrams and corresponding Feynman rules up to arbitrary order in (A)PT \cite{Bohm:2001yx,Maas:2017wzi,Maas:up}. Here we are explicitly interested in VBS in the scalar channel which additionally requires a partial wave analysis. This can be achieved by deconstructing the matrix according to
\begin{gather}\label{eqn:mat}
  \mathcal{M}     = 16\pi \sum_J (2J+1)f_J P_J(\cos\theta)\,,\\
  f_J             =\frac{1}{32\pi(2J+1)}\int_{-1}^{1} \mathcal{M}\,P_J(\cos\theta) \dd{\qty(\cos\theta)} = e^{\im\delta_J}\sin(\delta_J)
  \qqtext{and}
  \tan(\delta_J)  = \frac{f_J}{1+\im f_J}\,,\label{eqn:ps}
\end{gather}
with $f_J$ the partial transition amplitude, $P_J$ the Legendre polynomials and $\delta_J$ the phase shift. For VBS at the here investigated Born level it is additionally necessary to perform a reunitarization \cite{Kilian:2014zja,Jenny:2022atm}, requiring to replace the initial $f_J$ by $1/\qty(\Re(1/f_J)-\im)$.

From \cref{eqn:mat,eqn:ps} it can be seen that the phase shift $\delta_J$ in the respective partial wave fully characterizes the scattering process. Independent of the perturbative level, the finite extent of the particles is therefore going to modify the phase shift and the transition amplitude by
\begin{equation}
  \tan(\delta_J)\!\rightarrow \tan(\delta_J) - \tan(\Delta\delta_J)
  \qqtext{and}
  f_J\!\rightarrow f_J - \frac{\tan(\Delta\delta_J)}{\qty[\tan(\delta_J)+\im]\qty[\tan(\Delta\delta_J)-\tan(\delta_J)-\im]} = f_J - \Delta f_J
\end{equation}
respectively. The transition matrix is consequently also split into $\mathcal{M}\rightarrow \mathcal{M}_{f_J} -\mathcal{M}_{\Delta f_J}$. The ratio of the modified differential cross-section to the one obtained from (A)PT changes thus to
\begin{equation}
  \left.\qty(\frac{d\sigma}{d\Omega})_{\text{mod.}}\middle/\qty(\frac{d\sigma}{d\Omega})_{\text{PT}}\right. = \frac{\abs{\mathcal{M}_{f_J} - \mathcal{M}_{\Delta f_J}}^2}{\abs{\mathcal{M}_{f_J}}^2} = \abs{\frac{\qty(\mathcal{M}_{f_J} - \mathcal{M}_{\Delta f_J})^2}{\mathcal{M}_{f_J}^2}}
  = \abs{1-\frac{2\mathcal{M}_{\Delta f_J}}{\mathcal{M}_{f_J}}\left(1-\frac{\mathcal{M}_{\Delta f_J}}{2\mathcal{M}_{f_J}}\right)} \,.
\end{equation}
Finally, the influence of the finite extent close to the elastic threshold can be parameterized conveniently by introducing the scattering length $a_0$ \cite{Sakurai:2011zz}
\begin{equation}\label{eqn:scattering_length}
  \tan(\Delta \delta_J(s))\approx-a_0\sqrt{s-4m_W^2}\,,
\end{equation}
which is negative for a particle with finite extent and $a_0\ge 0$ for point-like objects.

\begin{figure}[t!]
  \centering
  \begin{subfigure}{0.32\textwidth}
    \includesvg[width=\linewidth]{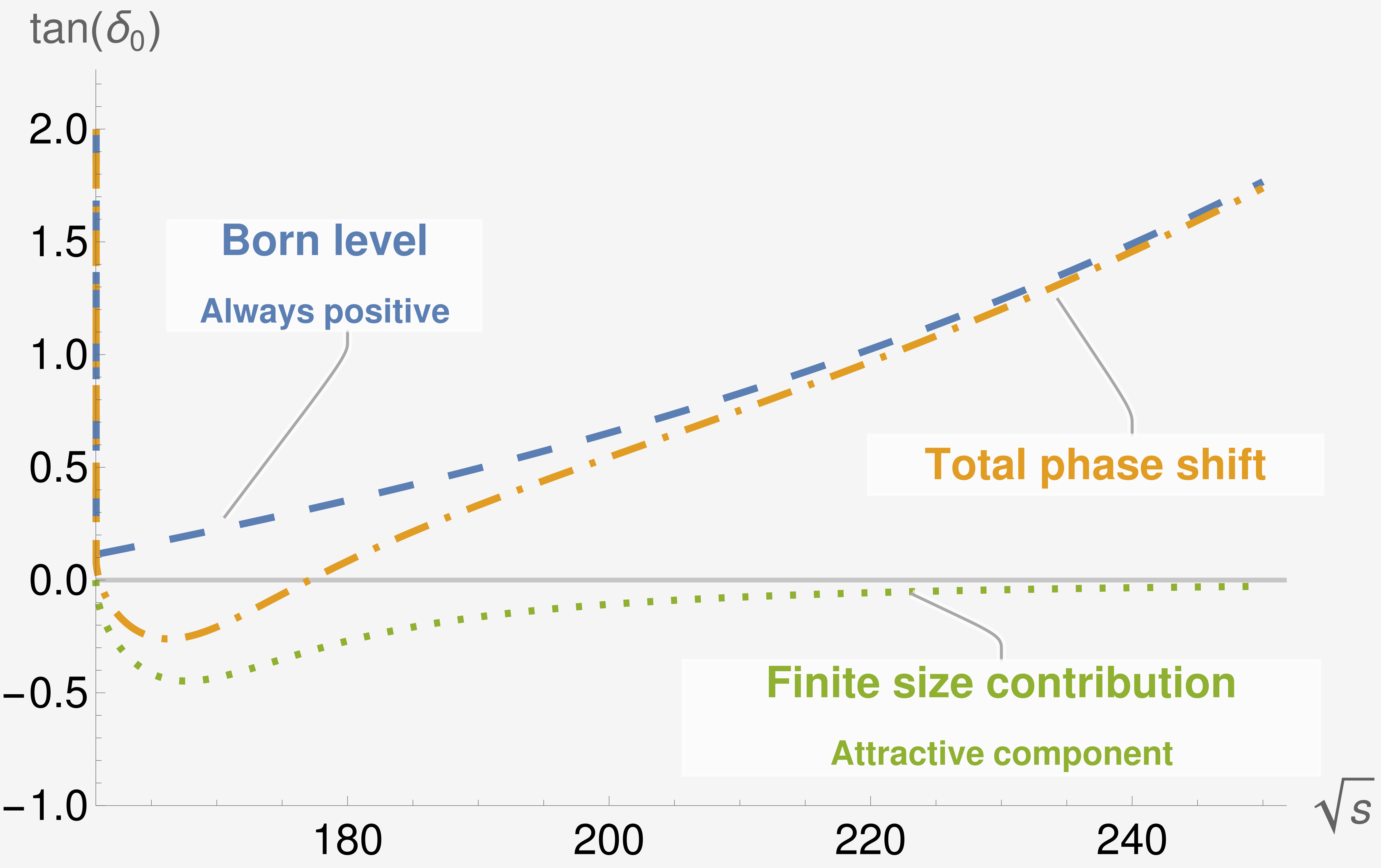}
    \caption{phase shift}
    \label{fig:ps_pt}
  \end{subfigure}
  \hfil
  \begin{subfigure}{0.32\textwidth}
    \includesvg[width=\linewidth]{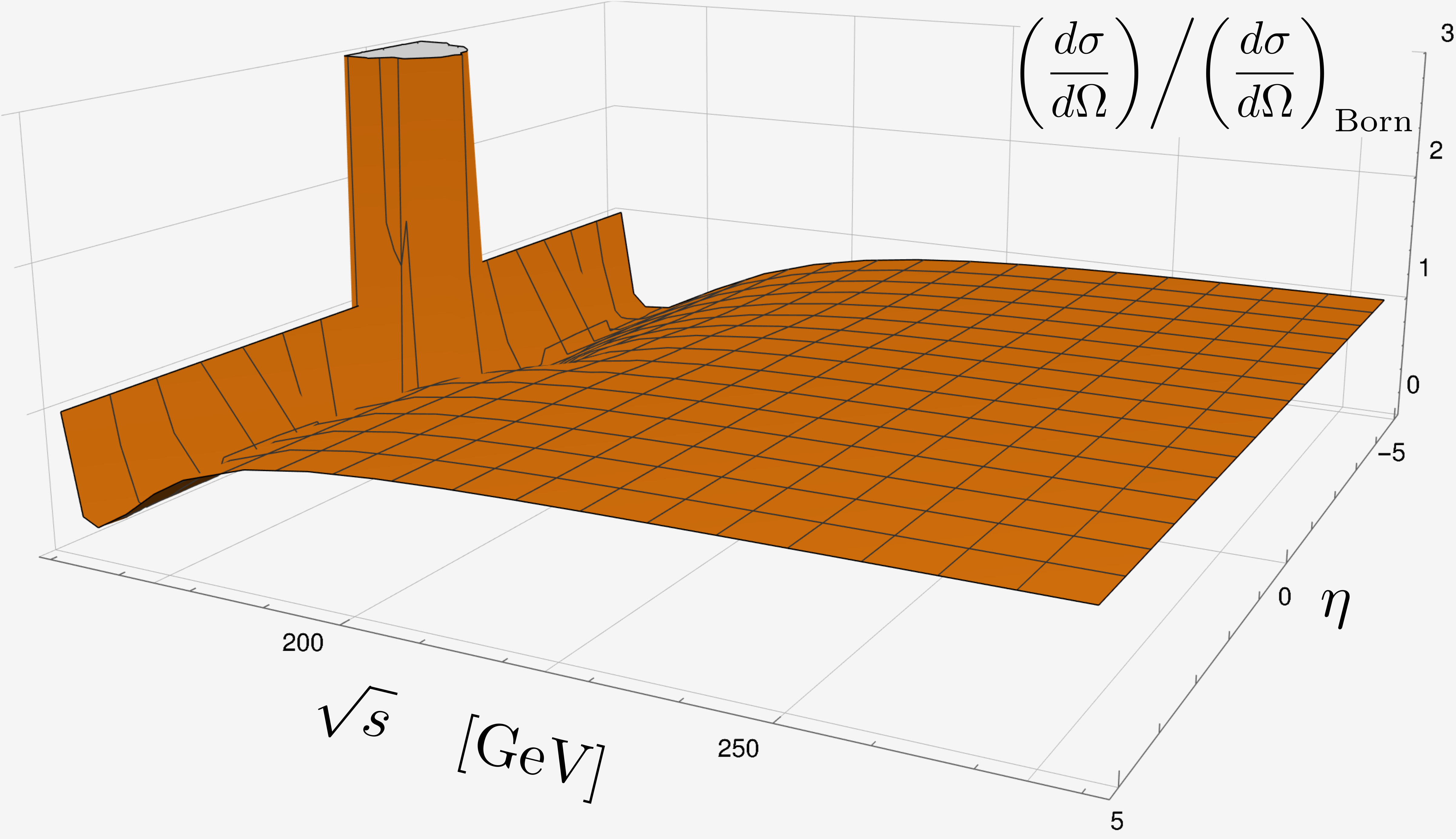}
    \caption{differential cross-section}
    \label{fig:dxs_pt}
  \end{subfigure}
  \hfil
  \begin{subfigure}{0.32\textwidth}
    \includesvg[width=\linewidth]{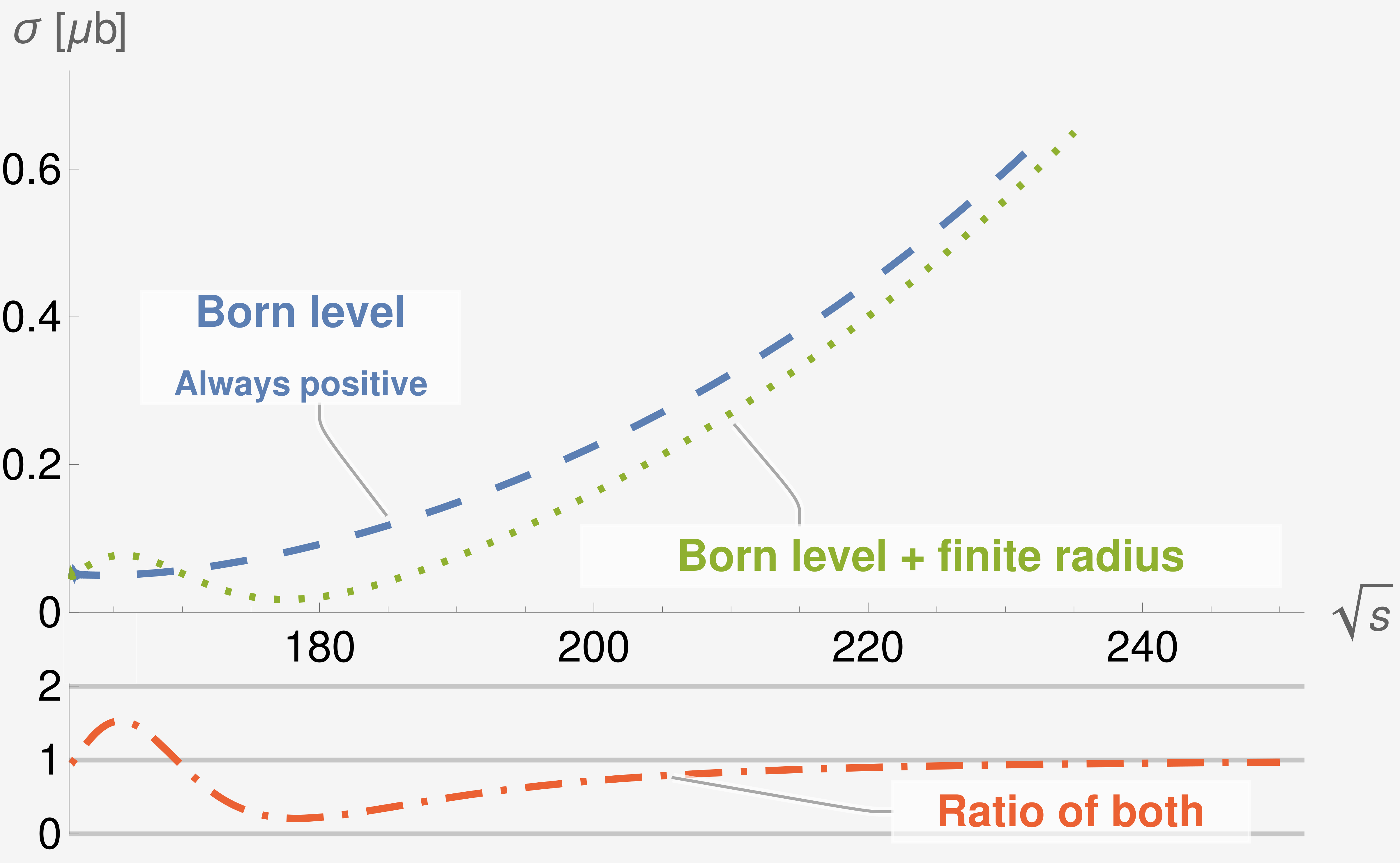}
    \caption{integrated cross-section}
    \label{fig:xs_pt}
  \end{subfigure}
  \caption{Expected modifications to VBS for scattering properties in the elastic region due to a Higgs with finite extent. Results obtained within a reduced SM-setup for a Higgs with mass $m_H=\SI{125}{\GeV}$ and scattering length $a_0^{-1}\approx-\SI{40}{\GeV}$.}
  \label{fig:augmented_pt}
\end{figure}

In \cref{fig:augmented_pt} we show the expected modifications of the different quantities for the VBS process with a finite-sized Higgs particle. The perturbative values have been obtained at Born-level for a reduced SM-setup with $m_W=m_Z=\SI{80.375}{\GeV}$ and $m_H=\SI{125}{\GeV}$. We see that modifications are only expected close to the threshold region and become negligible for higher momenta. Therefore, the differential and integrated cross-section in \cref{fig:dxs_pt,fig:xs_pt} respectively, show the typical profile for probing particles with some finite extend. We see suppressed forward- and backward scattering and enhanced scattering around zero rapidity $\eta$ for small center-of-mass energies. The results here are however considered to be only qualitative and not quantitative.

\section{Lattice techniques\protect\cref{note:skip}}\label{sec:lattice}

\begin{table}[ht!]
  \centering
  \caption{Parameter sets. Lattice spacing has been set by fixing the mass of the vector-boson to \SI{80.375}{\GeV} \cite{Maas:2013aia}. The running weak gauge coupling in the MiniMOM scheme \cite{vonSmekal:2009ae} has been determined as in \cite{Maas:2013aia}.}
  \label{tab:pars}
  \begin{tabular}{c|c|c|c|c|c|c}
    Name  & $\beta$ & $\kappa$ & $\gamma$ & $\alpha_{W,\SI{200}{\GeV}}$ & $a^{-1}\ \qty[\si{\GeV}]$ & $m_H\ \qty[\si{\GeV}]$ \\\hline\hline
    Set 1 & 4.0000  & 0.2850   & 0.970    & 0.219                       & 289                       & --                     \\
    Set 2 & 4.0000  & 0.3000   & 1.000    & 0.211                       & 243                       & $275^{+3}_{-3}$        \\
    Set 3 & 2.7984  & 0.2984   & 1.317    & 0.492                       & 287                       & $148^{+6}_{-20}$
  \end{tabular}
\end{table}

We study now the same process using lattice simulations, to compare it with the analytic expectations obtained in \cref{sec:higgs_radius} and to check the reliability of the FMS-approach. Therefore, we use the lattice discretization taken from \cite{Montvay:1994cy}. More details on the action and the simulation itself can be found in \cite{Maas:2013aia,Maas:2014pba,Riederer:2020tvj,Jenny:2022atm}. The lattice sizes used are $L^4$ with $L=\qty{8,12,\dots,32}$. In \cref{tab:pars} we show the parameters that will be further discussed in \cref{sec:results}. What should be mentioned however already at this point is that the bosonic degrees of freedom of this theory required us to employ very large statistics ($\order{10^{(4-6)}}$ configurations per set and volume) to overcome the huge signal-to-noise ratio (SNR). Additionally, we were also forced to simulate at very coarse lattices and relatively large weak-coupling compared to the SM to obtain the results in the following.

Obtaining the phase shift on the lattice in a specific scattering channel and partial wave \cite{Luscher:1990ux,Lang:2011mn,Briceno:2017max} can be straightforwardly applied to the weak sector. The procedure is a two-step process: First, obtaining the energy spectrum as a function of lattice-volume in the involved scattering channels, i.e.\ the $0^+$/scalar- and the $1^-$/vector-boson-channel, up to the inelastic threshold at $\min(4m_W,2m_H)$. Second, performing a Lüscher analysis \cite{Luscher:1990ux} which connects the volume-dependent energy levels on the lattice with the infinite volume phase-shift.

For the spectroscopic analysis we performed a variational analysis with an operator-basis of two vector- and 36 scalar-interpolators in the respective channels. The vector-interpolators are a direct discretization of (\ref{eqn:physical_ops}) with different normalizations. For the scalar-operators we also used the discretization of (\ref{eqn:physical_ops}) as well as a gauge-ball-interpolator, spatial-summed vector-interpolators, and combinations of the previous interpolators with finite back-to-back momenta. The full list can be found in \cite{Jenny:2022atm}. To reduce noise we employed APE smearing up to 4 times on all elementary fields. This leaves us with a full operator-basis of 10 vector- and 180 scalar-interpolators.

While extracting the energy spectrum with such a large basis in the scalar channel it turned out, that using only the largest smearing level improves the SNR significantly. To get the energy-levels from the correlators we used an alternative technique compared to the usual methods of fitting the effective energies \cite{Jenny:2022atm}. This method outperformed usual double-cosh fits by far. Finally, to find the best suited subset of operators with minimal SNR we performed an iterative procedure, where we subsequently increased the size of the variational basis depending on the time-averaged SNR of the individual correlators. This allowed us to identify of order 10--20 operators for each set and volume, and to obtain the energy spectra in the specific channels, with sufficient accuracy.

For the Lüscher analysis the lattice momenta $\va{p}$ have been obtained using the lattice dispersion relation $\cosh{\qty(E/2)} = \cosh{\qty(m_W)} + 2\sin{\qty(\va{p}/2)}^2$, with $m_W$ the obtained infinite volume mass in lattice units. The interactions will shift the lattice momenta away from the usual integer-multiples of $2\pi/L$ to some value $q =\abs{\va{p}} \frac{L}{2\pi}$. This can be used to extract information about the interaction by relating $q$ to the phase shift $\delta_J$ via a transcendental equation in terms of the so-called generalized zeta-function $\mathcal{Z}_{Jm}^{\va{d}}$ \cite{Luscher:1985dn,Luscher:1990ux}. For $J=0$ and vanishing center-of-mass momentum the defining equation is given by
\begin{equation}\label{eqn:phase shift}
  \tan(\delta_{0}\qty(q)) = \frac{\pi^{\frac{3}{2}}q}{\mathcal{Z}_{00}^{\va{0}}\qty(1,q^2)}\,.
\end{equation}

\section{Nonperturbative results}\label{sec:results}

The main obstacle when comparing results from (A)PT with lattice simulations is that all possible initial states on the lattice will mix. In the case of VBS this means that the scattering matrix needs to be constructed from all possible two $1^-$ states in an $s$-wave with zero total momentum and net-zero weak/custodial charge, i.e.\ $W^\pm W^\mp \rightarrow W^\pm W^\mp$, $W^\pm W^\mp \leftrightarrow ZZ$\footnote{Note that the $Z$ is degenerate with the $W^\pm$-bosons in the reduced SM-setup we are using.} and $ZZ \rightarrow ZZ$. This results in a sum over all 81 possible full 4-point vertices of the vector-operator from (\ref{eqn:physical_ops}) and finally yields a perturbative expression for the transition matrix that can be compared to the lattice results. The full expression is given in \cite{Jenny:2022atm}. Here, we show the non-perturbative results compared to the APT prediction at Born level for the three different parameter sets.

\begin{figure}[t!]
  \centering
  \begin{subfigure}{0.49\textwidth}
    \includegraphics[width=\linewidth]{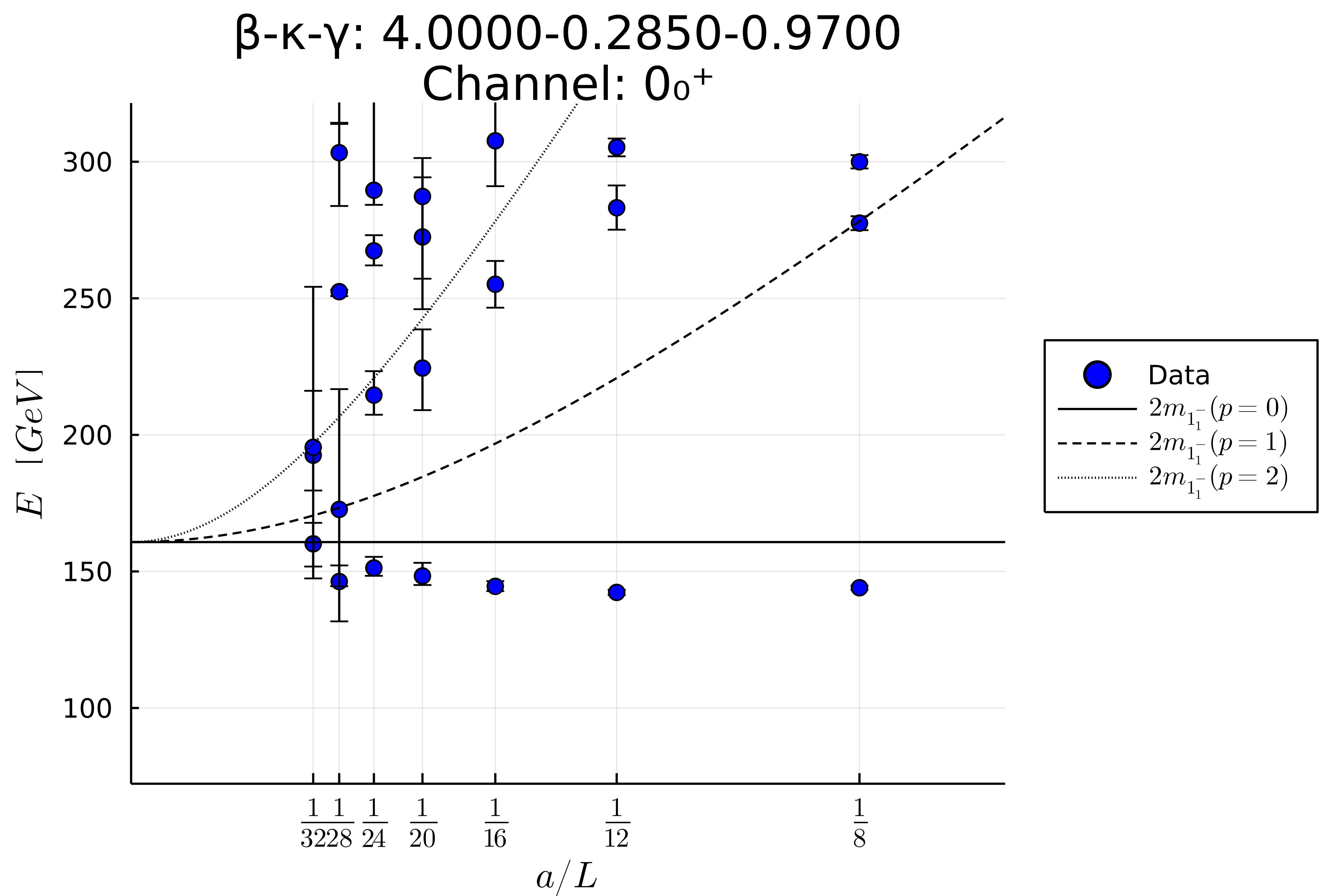}
    \caption{Set 1: energy spectrum}
    \label{fig:spectrum_heavy}
  \end{subfigure}
  \hfil
  \begin{subfigure}{0.49\textwidth}
    \includegraphics[width=\linewidth]{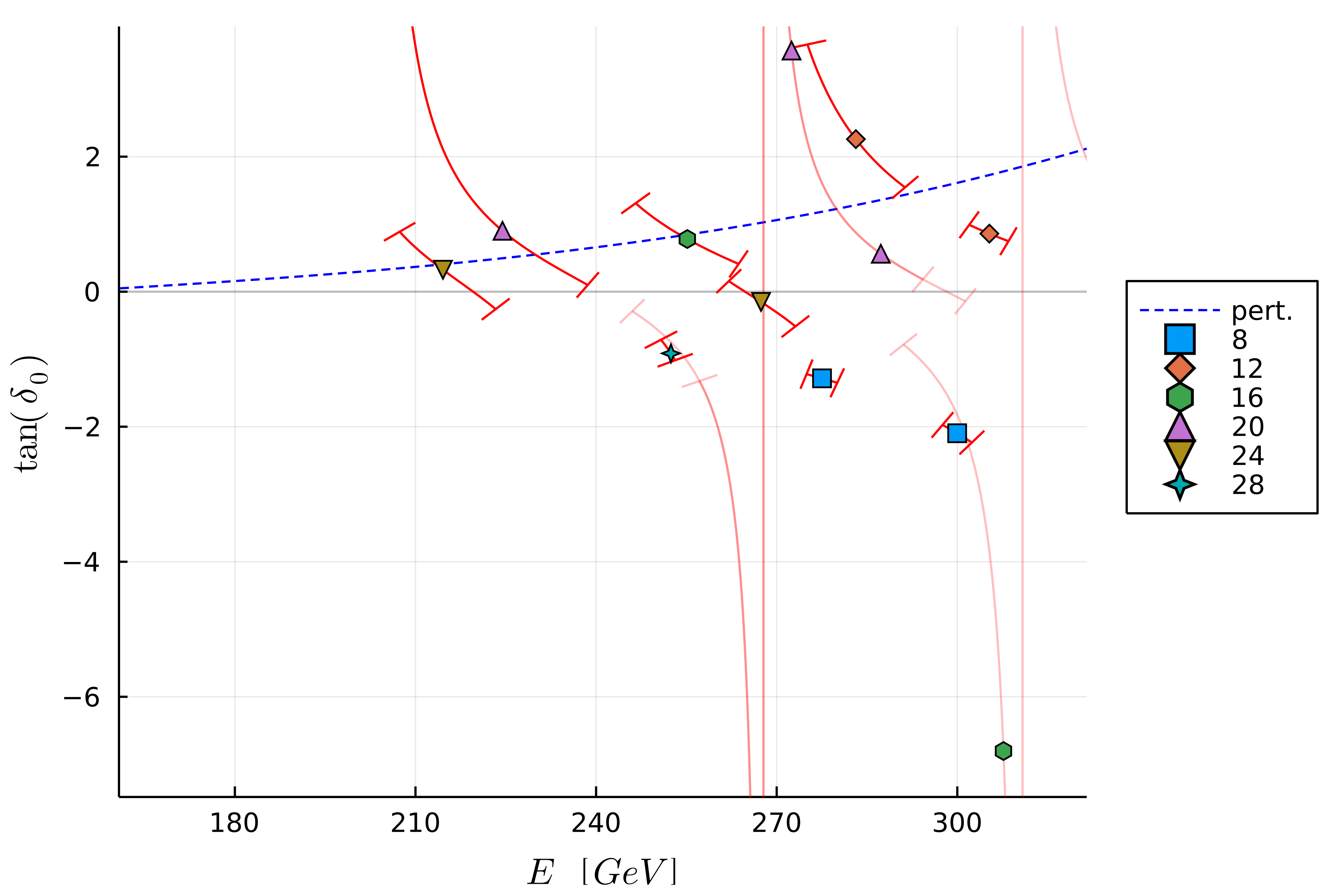}
    \caption{Set 1: phase shift}
    \label{fig:ps_heavy}
  \end{subfigure}\\
  \begin{subfigure}{0.49\textwidth}
    \includegraphics[width=\linewidth]{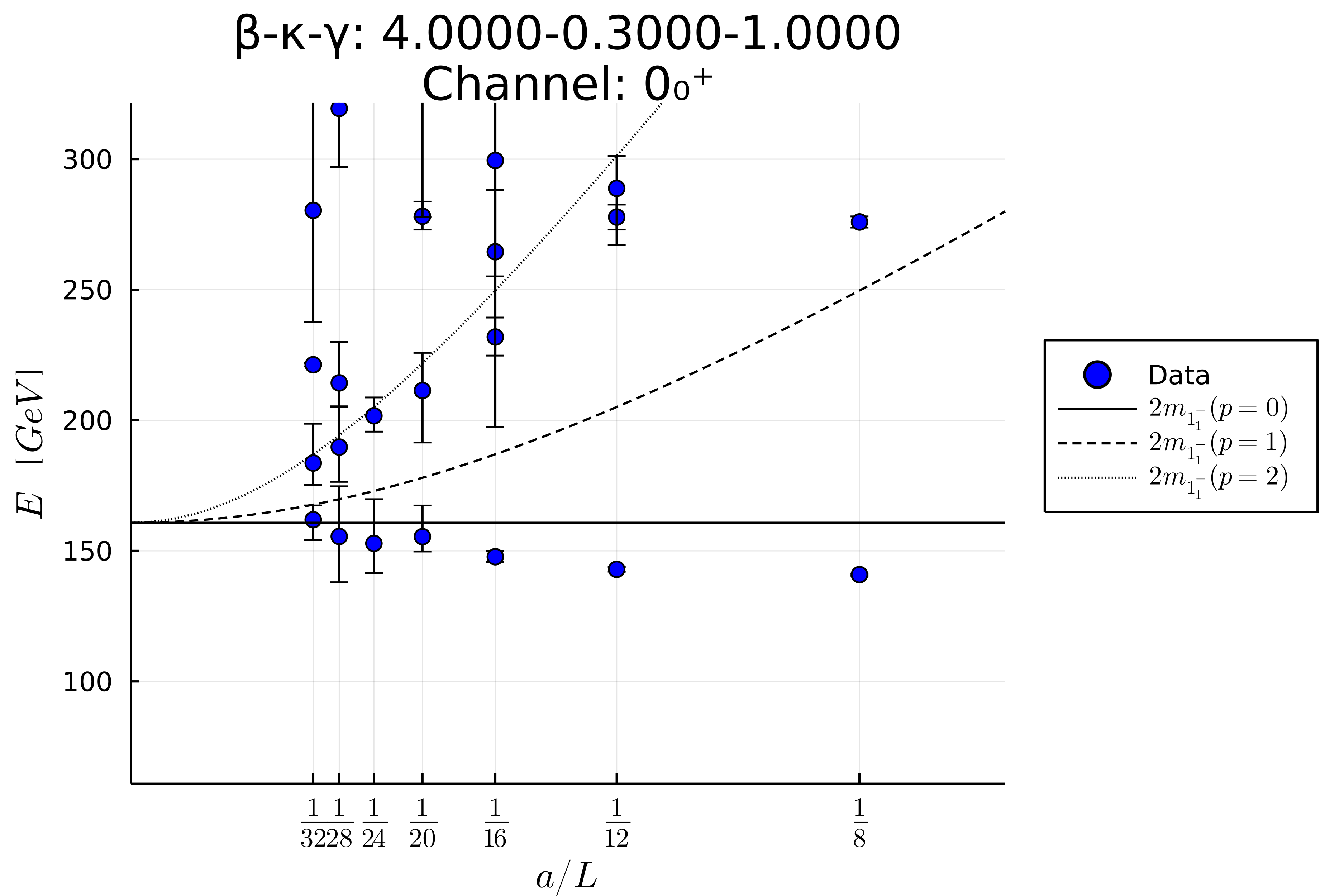}
    \caption{Set 2: energy spectrum}
    \label{fig:spectrum_resonance}
  \end{subfigure}
  \hfil
  \begin{subfigure}{0.49\textwidth}
    \includegraphics[width=\linewidth]{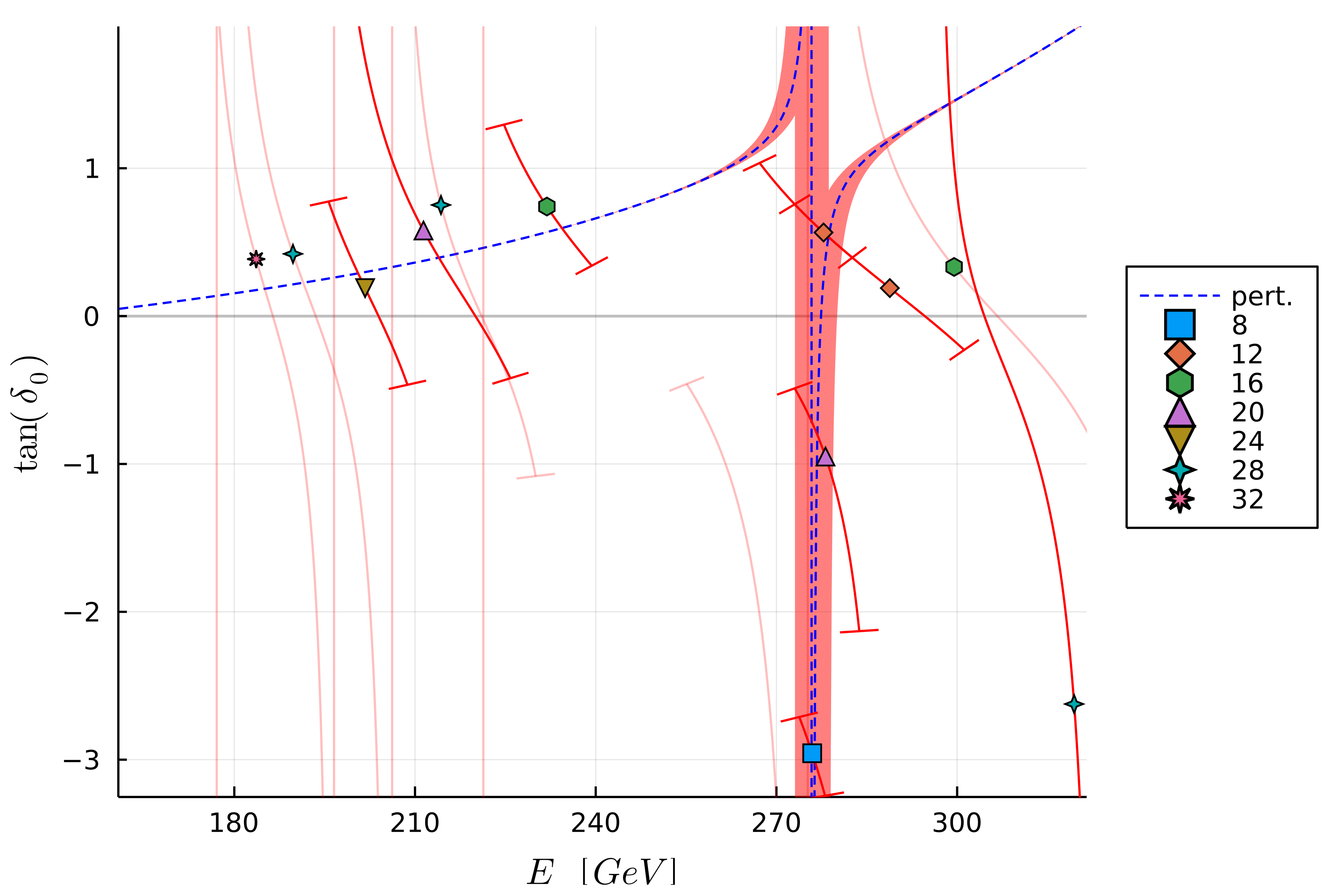}
    \caption{Set 2: phase shift}
    \label{fig:ps_resonance}
  \end{subfigure}
  \caption{Energy spectrum and expected states (lines) in the scalar channel against the inverse lattice size (left) and the corresponding phase shift (right) obtained from lattice simulations for set 1 and 2 (\cref{tab:pars}).}
  \label{fig:heavy_resonance}
\end{figure}

The simplest case is given in set 1 (\cref{tab:pars}), where we find massive vector-bosons but don't find a stable state below threshold in the scalar channel. This can be seen in \cref{fig:spectrum_heavy}. Since, the vector-bosons are massive, this means however that we are in a Higgs-like phase and would expect also a massive scalar-boson. Therefore, the two possibilities left are the Higgs-boson being either a resonance inside the elastic region or above the inelastic threshold. To settle this question we calculated the phase shift from the lattice data, which is shown in \cref{fig:ps_heavy}. Since we don't see any singularity like behavior here, which would be indicative of a resonance \cite{Briceno:2017max}, we can conclude that this set indeed describes a scenario with a Higgs-boson heavier than the inelastic threshold. This means that the diagrams containing a Higgs-propagator do not contribute to the VBS process. This is consistent with the prediction obtained from APT for this case.

In set 2 (\cref{tab:pars}) we see again a very similar picture for the energy spectrum (\cref{fig:spectrum_resonance}). However, when looking at the phase shift in \cref{fig:ps_resonance} we see this time a singularity like behavior around $\SI{275}{\GeV}$, being indicative of a resonance \cite{Briceno:2017max}. The usual approach of applying a Breit-Wigner fit to obtain the mass and the width, e.g.\ as in \cite{Lang:2011mn}, turned out to be non-reliable due to large uncertainties. On the other hand we also obtained an equation for the phase shift from APT with $m_H$ as a free parameter. Using this expression we performed a fit of the data and obtained the dashed curve in \cref{fig:ps_resonance}, giving a mass $m_H=\SI{275(3)}{\GeV}$. The obtained curve agrees quite well with the data which further supports the interpretation as a resonance in this channel. Though APT provides an alternative approach to obtaining resonances in this specific case, it should be mentioned that it is not possible to include a width of the resonance at Born level.

\begin{figure}[t!]
  \centering
  \begin{subfigure}{0.47\textwidth}
    \includegraphics[width=\linewidth]{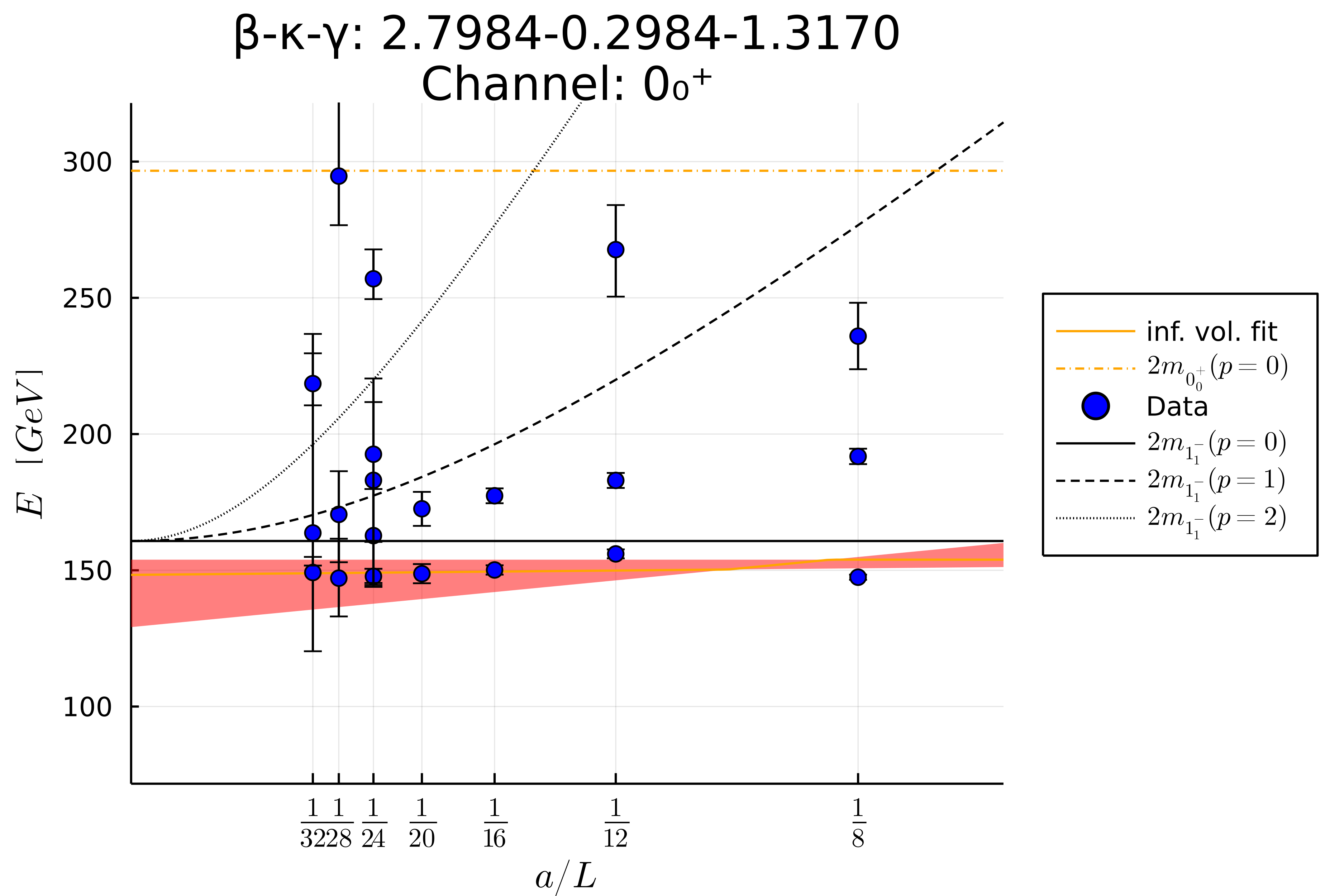}
    \caption{energy spectrum}
    \label{fig:spectrum_boundstate}
  \end{subfigure}
  \hfil
  \begin{subfigure}{0.47\textwidth}
    \includegraphics[width=\linewidth]{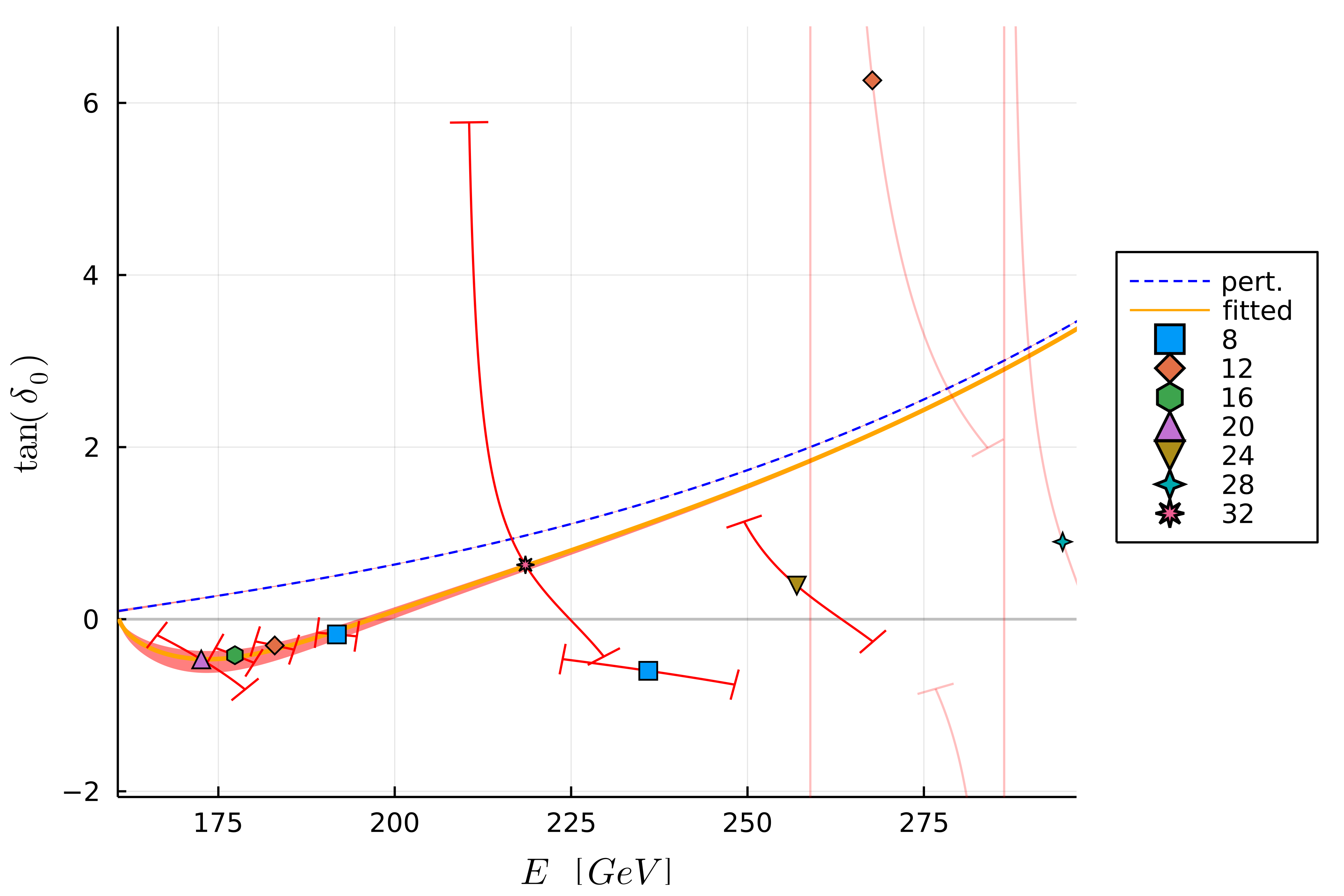}
    \caption{phase shift}
    \label{fig:ps_boundstate}
  \end{subfigure}\\
  \begin{subfigure}{0.47\textwidth}
    \includegraphics[width=\linewidth]{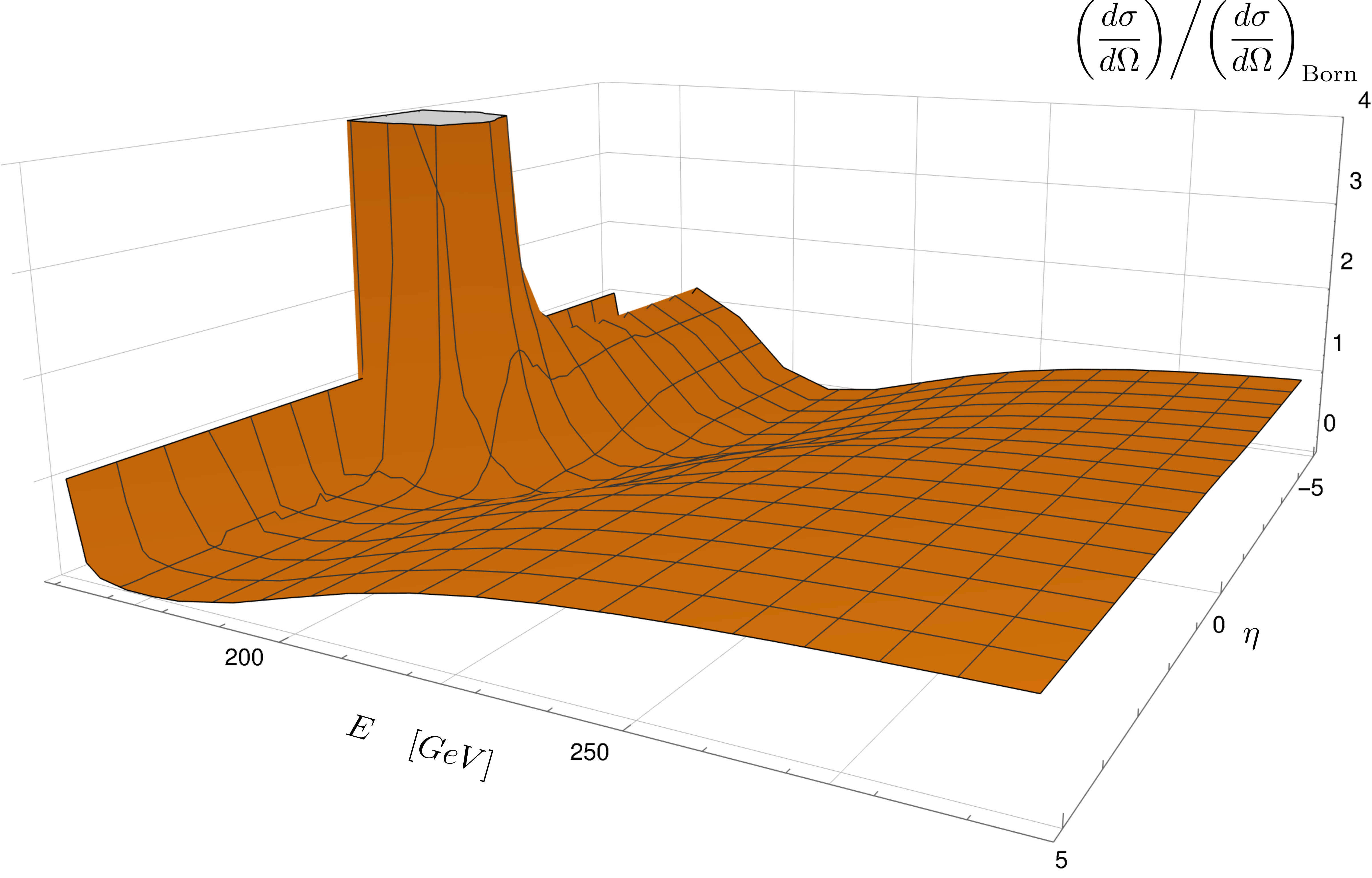}
    \caption{normalized differential cross-section}
    \label{fig:dxs_boundstate}
  \end{subfigure}
  \hfil
  \begin{subfigure}{0.47\textwidth}
    \includegraphics[width=\linewidth]{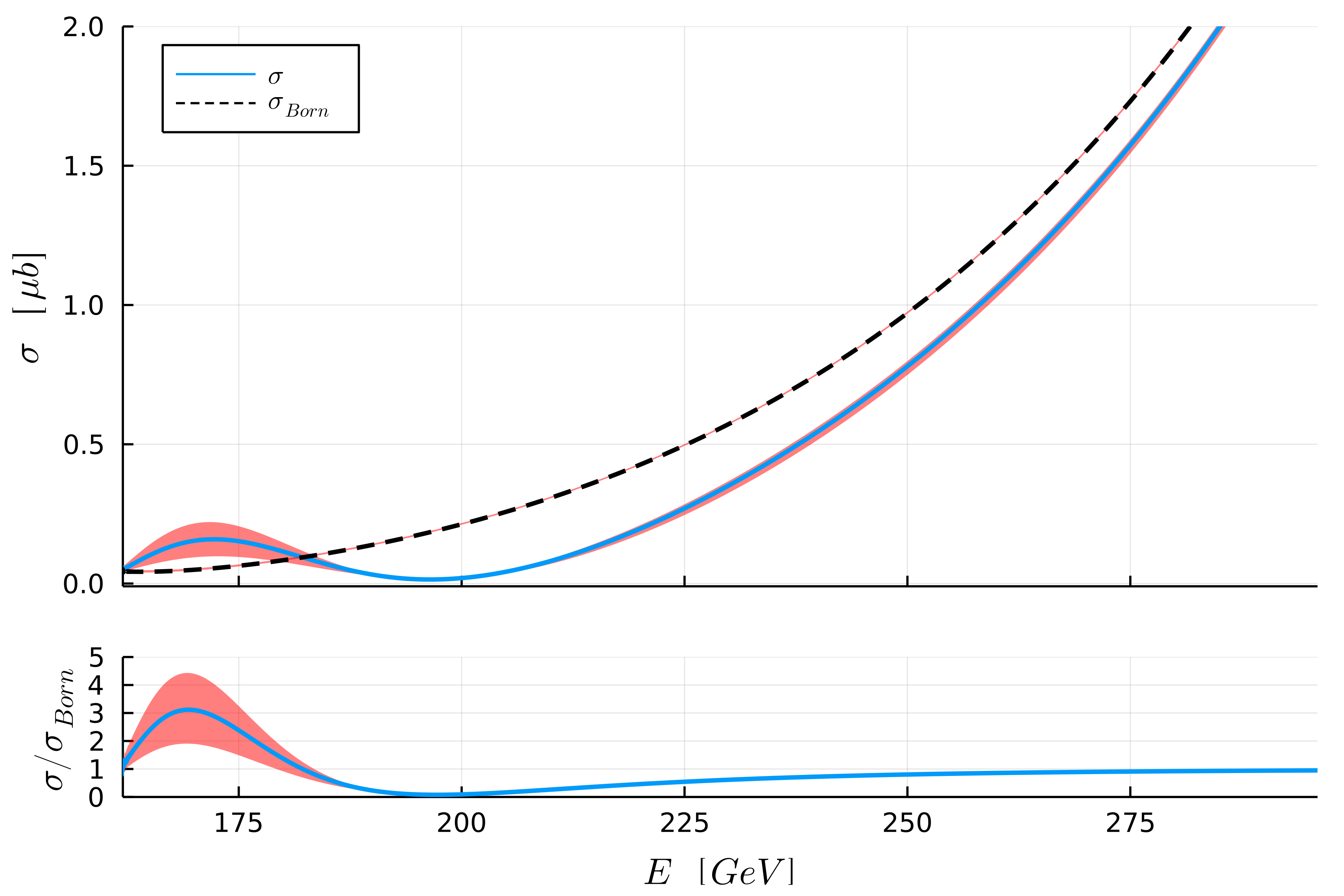}
    \caption{integrated cross-section}
    \label{fig:xs_boundstate}
  \end{subfigure}
  \caption{Results for set 3 (\cref{tab:pars}). Upper panel: Energy spectrum and expected states (lines) in the scalar channel against the inverse lattice size (left) and the corresponding phase shift (right). Lower panel: Normalized differential cross-section as a function of energy and pseudo-rapidity $\eta$ (left) and the corresponding integrated cross-section (right). The differential cross-section is normalized to the APT prediction.}
  \label{fig:boundstate}
\end{figure}

The last remaining possibility that needs to be explored is the one with a stable state in the scalar channel below the elastic threshold. This resembles the case of the SM and is found in set 3 (\cref{tab:pars}). The energy spectrum (\cref{fig:spectrum_boundstate}) shows an additional state below the elastic threshold, with an infinite volume mass of $m_H=148^{+6}_{-20}\si{\GeV}$. From the considerations in \cref{sec:higgs_radius} we would expect that this state is described by a particle with non-vanishing extent, and would result in a negative phase shift close to the threshold. In \cref{fig:ps_boundstate} we see this expected behavior with the data lying significantly and consistently below zero close to the threshold. In addition, the naive perturbative prediction does not agree with the data at all here. Therefore, we used the method as described in \cref{sec:higgs_radius} to include the finite extent of the Higgs-boson\footnote{Here we included also a second order term in \cref{eqn:scattering_length} to get better agreement throughout the elastic region \cite{Jenny:2022atm}.} and obtained a scattering length of roughly $-\SI{40}{\GeV}$. Remarkably this result is in agreement with a previous investigation of the weak radius for the physical vector bosons \cite{Maas:2018ska}, although using completely different techniques.

In \cref{fig:dxs_boundstate,fig:xs_boundstate} we show the differential and integrated cross-section obtained from the lattice simulations. Here we see exactly the picture that we have predicted in \cref{sec:higgs_radius} for a Higgs with a finite radius\footnote{The radius is not described by APT at Born level, which is also not expected. Whether it is a genuine non-perturbative effect or whether it is captured by APT beyond Born level remains to be seen.}. We therefore see, that lattice simulations indeed support the elementary degree of freedom in the scalar channel being a bound state operator as in \cref{eqn:physical_ops} rather than the elementary field itself. Therefore, it is possible to study this from deviations of VBS cross-sections at experiments, at least in principle.

\section{Conclusions and outlook}

We have presented a fully gauge-invariant study of the VBS process. To get a conclusive picture we tackled the problem with two different approaches once using augmented perturbation theory (APT) and once using lattice simulations.

The gauge-invariant approach requires using manifestly gauge-invariant operators as initial and final states rather than elementary fields, as is usually done in PT. However, this additionally leads to the asymptotic states being bound-states with a non-vanishing extent. Therefore, this allows us to simultaneously probe the theoretical foundations of the standard model, as well as deriving a description for VBS with any kind of bound-states involved, like e.g.\ a composite Higgs from BSM. We showed that the finite extent modifies VBS close to threshold as expected for bound-state scattering and can thus be used as a possible avenue for experiments to search for new physics.

The non-perturbative approach also supports the previous statements in the case where a stable state is found below the elastic threshold. Here we have seen a significant deviation of the data from the usual perturbative prediction. By including the finite extent of the Higgs boson we were again able to compensate the discrepancy and find again the expected behavior for VBS.

Additionally\cref{note:skip}, the lattice also provided a possibility to study cases with a Higgs boson heavier than the elastic threshold. We found that in this case the predicted phase shifts from APT are in good agreement with the data, which suggests that non-perturbative effects or higher-order contributions are small in this case. Nevertheless, especially in the case of a resonance, APT could be used as a tool to extract resonance properties which has not been possible by standard methods. This provides therefore an alternative in theories with a BEH effect due to the FMS-mechanism and APT.

\acknowledgments

B. R. has been supported by the Austrian Science Fund FWF, grant P32760. The computational results presented have been obtained using the HPC center at the University of Graz. We are grateful to its team for its exceedingly smooth operation.

\bibliographystyle{JHEP}
\bibliography{refs}

\end{document}